# Throughput Scaling Laws for Wireless Networks with Fading Channels

Masoud Ebrahimi, Mohammad Maddah-Ali, and Amir Khandani
Electrical and Computer Engineering Department
University of Waterloo, Waterloo, ON, Canada, N2L 3G1
Emails: {masoud, mohammad, khandani}@cst.uwaterloo.ca

*Abstract*—A network of $n$ wireless communication links is considered. Fading is assumed to be the dominant factor affecting the strength of the channels between nodes. The objective is to analyze the achievable throughput of the network when power allocation is allowed. By proposing a decentralized on-off power allocation strategy, a lower bound on the achievable throughput is obtained for a general fading model. In particular, under Rayleigh fading conditions the achieved sum-rate is of order $\log n$, which is, by a constant factor, larger than what is obtained with a centralized scheme in the work of Gowaikar et al. Similar to most of previous works on large networks, the proposed scheme assigns a vanishingly small rate for each link. However, it is shown that by allowing the sum-rate to decrease by a factor $\alpha < 1$, this scheme is capable of providing non-zero rate-per-links of order $\Theta(1)$. To obtain larger non-zero rate-per-links, the proposed scheme is modified to a centralized version. It turns out that for the same number of active links the centralized scheme achieves a much larger rate-per-link. Moreover, at large values of rate-per-link, it achieves a sum-rate close to $\log n$, i.e., the maximum achieved by the decentralized scheme.

## I. INTRODUCTION

In a wireless network, a number of source nodes transmit data to their designated destination nodes through a shared wireless channel. Followed by the pioneering work of Gupta and Kumar [1], considerable attention has been paid to investigate how the throughput of wireless networks scales with $n$, the number of nodes. This has been done assuming different network topologies, traffic patterns, protocol schemes, and channel models [1]–[11].

Most of the works analyzing the throughput of wireless networks consider a channel model in which the signal power decays according to a distance-based attenuation law [1], [3]–[10]. However, in a wireless environment the presence of obstacles and scatterers adds some randomness to the received signal. This random behaviour of the channel, known as fading, can drastically change the scaling laws of a network [2], [11].

A common characteristic of [1]–[11] is their use of *multihop* communications, which necessitates the presence of a central unit with full knowledge of all channel conditions to decide on the routing paths and the schedule of transmissions. Furthermore, multihop communications induces delay in the network and excess power consumption by the relay nodes.

In this paper, we follow the model of [11], where fading is assumed to be the dominant factor affecting the strength of the channels between nodes. The network is limited to single-hop communications to avoid the complexity and other drawbacks of multihop communications mentioned above. We propose a link activation strategy that can be implemented in a decentralized fashion. The only information required at each transmitter is its direct channel coefficient. Analysis of the proposed strategy shows that the throughput of the network strongly depends on the fading channel distribution. In particular, in a Rayleigh fading channel, it achieves a throughput that scales like $\log(n)$; this is larger (by a constant factor) than what is achieved in [11] by a complicated centralized scheme.

In most of previous works, e.g. [1]–[4], [11], rate-per-links decrease to zero as the number of nodes grows. However, it is practically appealing to have rate-per-links that remain constant when $n$ grows. In [9], it is shown that a nondecreasing rate per node is achievable when nodes are mobile. In this work, we show that our proposed strategy is capable of providing rate-per-links like $\Theta(1)$ while keeping the order of the maximum sum-rate unchanged up to a constant factor. To improve the proposed scheme, we modify it by adding an interference management phase to it. This new scheme should be implemented in a centralized fashion.

The rest of the paper is organized as follows. In Section II, the network model and objective are described. We derive a lower bound on the sum-rate in Section III. The analysis of the proposed method for some specific fading models is presented in Section IV. Finally, we address the issue of fixed rate-per-link in Section V and show how to improve it.

*Notation:* log is the natural logarithm function; P($A$) denotes the probability of event $A$; E($x$) represents the expected value of the random variable $x$; $\approx$ means approximate equality; for any functions $f(n)$ and $h(n)$, $h(n) = o(f(n))$ is equivalent to $\lim_{n\to\infty} |h(n)/f(n)| = 0$, $h(n) = \omega(f(n))$ is equivalent to $\lim_{n\to\infty} |h(n)/f(n)| = \infty$, $h(n) = \Theta(f(n))$ is equivalent to $\lim_{n\to\infty} |h(n)/f(n)| = c$, where $0 < c < \infty$, and $h(n) \sim f(n)$ is equivalent to $\lim_{n\to\infty} h(n)/f(n) = 1$; an event $A_n$ holds *asymptotically almost surely* (a.a.s) if P($A_n$) $\to 1$ as $n \to \infty$.

## II. NETWORK MODEL

We consider a wireless communication network with $n$ pairs of transmitters and receivers. Each transmitter aims to send data to its corresponding receiver. The transmit power of link $i$ is denoted by $p_i$ and is constrained as $0 \leq p_i \leq P$, where $P > 0$ is the maximum allowed transmit power.

The channel between transmitter $j$ and receiver $i$ is represented by a coefficient $g_{ji}$. This means the received power from transmitter $j$ at the receiver $i$ equals $g_{ji}p_j$. We refer to coefficients $g_{ii}$ and $g_{ji}$ ($j \neq i$) as *direct channel* and *cross channel coefficients*, respectively. In this paper, we assume the channel coefficients are i.i.d. random variables drawn from a pdf $f(x)$ with mean $\mu$ and variance $\sigma^2$.

We consider an additive white Gaussian noise (AWGN) with limited variance $\eta$ at the receiver $i$. The receivers are conventional, linear receivers, i.e., without multiuser detection. Since the transmissions occur simultaneously within the same environment, the signal from each transmitter acts as interference for other links. Assuming Gaussian signal transmission from all links, the distribution of the interference will be Gaussian as well. Thus, the maximum supportable rate of link $i$ is obtained as

$$r_i = \log\left(1 + \frac{g_{ii}p_i}{\eta + \sum_{\substack{j=1 \\ j \neq i}}^{n} g_{ji}p_j}\right) \quad \text{nats/channel use.} \quad (1)$$

The throughput (or sum-rate) of the network is defined as

$$R = \sum_{i=1}^{n} r_i. \quad (2)$$

In this paper, we use the terms throughput and sum-rate interchangeably. We desire to choose the transmission powers $p_i$ such that $R$ is maximized. However, the problem of throughput maximization is not convex and its optimum solution cannot be obtained easily. Instead, one can utilize suboptimum power allocation schemes to obtain lower bounds on the maximum achievable throughput.

Motivated by the low *SINR* (Signal to Interference-plus-Noise Ratio) regime in which the optimum powers are either zero or the maximum allowed value [12], we base our discussions in this paper on *on-off power allocation strategies*.

*Definition 1:* A power allocation is called an *on-off strategy* if the power of link $i$ is selected from the set $\{0, P\}$.

### III. A LOWER BOUND ON SUM-RATE

Consider the following on-off power allocation strategy.

*Strategy 1:* For a threshold $t$, choose the transmit power of link $i$ as

$$p_i = \begin{cases} P & ; \quad g_{ii} > t \\ 0 & ; \quad g_{ii} \leq t \end{cases}. \quad (3)$$

In this strategy, the quality of the direct channel of each link determines whether that link is active or silent. If each transmitter is aware of its direct channel coefficient, it can individually determine its transmit power. Hence, Strategy 1 can be implemented in a *decentralized* fashion. The sum-rate achieved by Strategy 1 provides a lower bound to the sum-rate of the wireless network under consideration.

The performance of Strategy 1 depends on the value of the threshold $t$; if $t$ is very large, the quality of the selected links will be very good, but the number of such links is small and as a result the achieved sum-rate will be small; on the other hand, if $t$ is very small, many links are chosen, but it causes a large interference and again the sum-rate will be small. Thus, it is crucial to choose a proper value for $t$.

Let $k$ denote the number of links activated under Strategy 1. Without loss of generality, we assume that the active links are indexed by $1, 2, \cdots, k$. From (1), (2), and (3), the corresponding sum-rate is equal to

$$R = \sum_{i=1}^{k} \log\left(1 + \frac{g_{ii}}{\rho + I_i}\right),$$

where $\rho = \eta/P$ and $I_i = \sum_{\substack{j=1 \\ j \neq i}}^{k} g_{ji}$ is the interference seen by link $i$. Considering the fact that $g_{ii} > t$ for the selected links and by using the Jensen's inequality, we obtain the following lower bound on the sum-rate

$$R \geq k \log\left(1 + \frac{t}{\rho + \frac{1}{k}\sum_{i=1}^{k} I_i}\right). \quad (4)$$

To make this lower bound more tractable, we use the Chebyshev inequality to write (4) as (see [12] for details)

$$R \geq k \log\left(1 + \frac{t}{\mu k + \psi_k}\right) \quad a.a.s. \quad (5)$$

for any $\psi_k \to \infty$ as $k \to \infty$.

Assume the probability of a link being active is $q$. Due to Strategy 1, which selects links independently and with probability $q$, the number of active links, $k$, is a binomial random variable with parameters $n$ and $q$.

*Lemma 1:* If $k$ is a binomial random variable with parameters $n$ and $q$, then for any $\xi_n \to \infty$ as $n \to \infty$ we have

$$|k - nq| < \xi_n \sqrt{nq} \quad a.a.s.$$

*Proof:* See [12]. ■

Now, we are ready to prove the main result of this section, which is an achievability result on the sum-rate.

*Theorem 1:* Consider a wireless network with $n$ links and i.i.d. random channel coefficients with pdf $f(x)$, cdf $F(x)$, and mean $\mu$. Choose any $t > 0$ and define $q = 1 - F(t)$. Then, a sum-rate of

$$R(t) = (nq - \xi_n\sqrt{nq})\log\left(1 + \frac{t}{\mu(nq - \xi_n\sqrt{nq}) + \varphi_n}\right) \quad (6)$$

is a.a.s. achievable for any $\xi_n = o(\sqrt{nq})$ that approaches infinity as $n \to \infty$ and any $\varphi_n = \psi(nq - \xi_n\sqrt{nq})$, where $\psi(n)$ makes the function

$$n \log\left(1 + \frac{c}{\mu n + \psi(n)}\right) \quad (7)$$

increasing in $k$ for any constant $c$, and $\psi(n) \to \infty$ as $n \to \infty$.

*Proof:* By applying Strategy 1, we obtained a lower bound on the achievable sum-rate in (5). According to condition (7), this lower bound is increasing in $k$. Thus, if we replace the number of active links, $k$, by its possible lower bound, the achievability result remains valid. However, from Lemma 1 we know that

$$k > nq - \xi_n\sqrt{nq} \quad a.a.s. \quad (8)$$

By the assumption $\xi_n = o(\sqrt{nq})$, the lower bound in (8) is non-negative and can replace $k$ in (5) to give (6) with $\varphi_n = \psi(nq - \xi_n\sqrt{nq})$. ∎

To facilitate working with the result of Theorem 1, in the sequel we will ignore the terms containing $\xi_n$ and $\psi_n$. This is justified by the fact that these functions can be chosen arbitrarily small and we are only interested on the order of parameters such as $R$, $t$, and $k$. Thus, we have the following simplified corollary.

*Corollary 1:* Under the conditions described in Theorem 1, the achieves the sum-rate and the number of active links in Strategy 1 are obtained as

$$R(t) = nq \log\left(1 + \frac{t}{\mu nq}\right) \quad a.a.s. \quad (9)$$
$$k \sim nq \quad a.a.s. \quad (10)$$

By defining the rate-per-link as $\lambda = R/k$, we have the following corollary.

*Corollary 2:* The rate-per-link in Strategy 1 scales as

$$\lambda \sim \log\left(1 + \frac{t}{\mu nq}\right) \quad a.a.s.$$

As specified in (6), the achievable sum-rate in Theorem 1 is a function of the parameter $t$. Thus, $t$ can be chosen such that the achievable sum-rate in Theorem 1 is maximized. Let's define

$$t^* = \arg\max_t R(t),$$
$$R^* = \max_t R(t)$$

to be the optimum threshold and the maximum achievable sum-rate, respectively. In general, the values of $t^*$ and $R^*$ are functions of $n$, but how they scale with $n$ strongly depends on the channel distribution function $f(x)$. This will be elaborated on in the next section through some examples of fading channels.

## IV. CASE STUDY

In this section, we apply the results of the previous section to the Rayleigh and log-normal fading channel models. It turns out that the sum-rate achieved by our decentralized scheme is larger than what is obtained in [11] for both cases. The proofs of the results can be found in [12].

### A. Rayleigh Fading

In a Rayleigh fading channel, the coefficients $g_{ji}$ are exponentially distributed with $f(x) = e^{-x}$ and $\mu = 1$. Thus, the relation between $q$ and $t$ is described as $q = e^{-t}$. By substituting this value in (9) and (10), we obtain the sum-rate and the number of active users as

$$R(t) = ne^{-t} \log\left(1 + \frac{t}{ne^{-t}}\right) \quad a.a.s. \quad (11)$$
$$k \sim ne^{-t} \quad a.a.s. \quad (12)$$

It can be verified that the optimum threshold that maximizes (11) is equal to

$$t^* = \log n - 2\log\log n + \log 2 + o(1).$$

*Corollary 3:* In a wireless network with $n$ links and under Rayleigh fading channel model, Strategy 1 with optimum threshold $t^*$ yields

$$R^* = \log n - 2\log\log n + O(1) \quad a.a.s.$$
$$k = \frac{1}{2}\log^2 n(1 + o(1)) \quad a.a.s.$$
$$\lambda \sim \frac{2}{\log n} \quad a.a.s.$$

The achieved sum-rate is at least by a factor of four larger than what is obtained in the centralized scheme of [11].

### B. Log-normal Fading

Consider a network with channel strengths drawn i.i.d. from a log-normal distribution with parameters $S$ and $M$ [11]. It can be easily verified that for large values of $t$, the following relation holds between $q$ and $t$

$$q \approx \frac{S}{\sqrt{2\pi}(\log t - M)} \exp\left(-\frac{(\log t - M)^2}{2S^2}\right).$$

By substituting this value of $q$ in (9), the sum-rate, as a function of $t$, is obtained as

$$R(t) = \frac{S}{\sqrt{2\pi}} \frac{n}{u} e^{-\frac{u^2}{2S^2}} \log\left(1 + \frac{Bue^u e^{\frac{u^2}{2S^2}}}{n}\right), \quad (13)$$

where, for the brevity of notation, we have defined $u = \log t - M$. Also, $B$ is a constant depending on the distribution parameters. It can be verified that the maximizing $t$ for (13) is equal to

$$t^* \sim e^{M-S^2} e^{S\sqrt{2\log n}}.$$

*Corollary 4:* In a wireless network with $n$ links and under log-normal fading channel model, Strategy 1 with optimum threshold $t^*$ yields

$$R^* \sim e^{-\frac{3S^2}{2}} e^{S\sqrt{2\log n}} \quad a.a.s.$$
$$k \sim \frac{e^{-\frac{3S^2}{2}}}{\sqrt{8}S} \sqrt{\log n} e^{S\sqrt{2\log n}} \quad a.a.s.$$
$$\lambda \sim \frac{\sqrt{8}S}{\sqrt{\log n}} \quad a.a.s.$$

The sum-rate scaling of $e^{S\sqrt{2\log n}}$ is, at least by a factor of $\log n / \log\log n$, larger than what is obtained in the centralized scheme of [11].

## V. NON-ZERO RATE-PER-LINK

In the previous section, it was shown that under Rayleigh fading model[1] Strategy 1 provides rate-per-links that scale as $\frac{2}{\log n}$. This rate, that approaches zero as $n \to \infty$, is not desirable in a practical scenario where each link requires a constant rate in the order of $\Theta(1)$. In this section, we show that the on-off strategy is capable of providing such amounts of rate-per-link. First, we need the following lemma that expresses the achievable sum-rate in terms of the number of active links.

---
[1]For brevity, in this section we only study the Rayleigh fading model. However, the same arguments can be presented for other fading models.

*Lemma 2:* In a wireless network with Rayleigh fading channels, if the $k$ best links (links with largest direct channel coefficients) are permitted transmission, then, the a.a.s. achievable sum-rate is related to $k$ according to (14) at the bottom of this page.

*Proof:* The proof relies on (11) and (12) that relate the parameters $t$, $k$, and $R$. We rewrite (12) in the following form, which is more useful in this proof:

$$t \sim \log n - \log k.$$

In what follows, the proof is presented in a case by case basis.

1) If $k = o(\log n)$, we can write $k = \dfrac{\log n}{\zeta_n}$, where $\zeta_n \to \infty$ as $n \to \infty$. Hence, we obtain

$$t = \log n - \log \log n + \log(\zeta_n).$$

Consequently, from (11) we obtain

$$R \sim \frac{\log n}{\zeta_n} \log \left( \frac{\log n}{\frac{\log n}{\zeta_n}} \right) = o(\log n).$$

2) If $k \sim \alpha \log n$ for some constant $\alpha > 0$, we have

$$t \sim \log n - \log \log n - \log \alpha.$$

As a result, we obtain

$$R \sim \alpha \log n \log \left( 1 + \frac{\log n}{\alpha \log n} \right)$$
$$= \alpha \log \left( 1 + \frac{1}{\alpha} \right) \log n$$

3) If $k = \omega(\log n)$ and $k = o(n^\alpha)$ for all $0 < \alpha < 1$, we should have $t \sim \log n$. Hence, we obtain

$$R \sim \omega(\log n) \log \left( 1 + \frac{\log n}{\omega(\log n)} \right) \sim \log n,$$

where the last equality is based on the approximation $\log(1+x) \approx x$ for small values of $x$.

4) If $k \sim n^{\alpha+\epsilon_n}$ for some $0 < \alpha < 1$ and some $\epsilon_n \to 0$, we should have

$$t \sim \log n - \log n^{\alpha+\epsilon_n} \sim (1-\alpha) \log n.$$

Substituting these values of $k$ and $t$ in (11) gives

$$R \sim (1-\alpha) \log n.$$

5) If $k \sim n^{1-\epsilon_n}$ for some $\epsilon_n \to 0^+$, we have

$$t \sim \log n - \log n^{1-\epsilon_n} = \epsilon_n \log n = o(\log n),$$

which results in $R = o(\log n)$. ∎

Each line in (14) describes a range of variations for $k$ that leads to the specified order of sum-rate. For example, the third line corresponds to the maximum sum-rate, $\log n$. In this section, we are interested in two other ranges of $k$ that lead to sum-rates of order $\Theta(\log n)$. These ranges include $k \sim \alpha \log n$ and $k \sim n^{\alpha+\epsilon_n}$ and will be addressed in the following subsections separately.

### A. Non-Zero Rate-per-Links by Decentralized On-Off Strategy

According to Lemma 2, when $k = \Theta(\log n)$, sum-rate is of order $\Theta(\log n)$ as well. In this case, rate per link scales as $\Theta(1)$. More precisely, in this range of operation we have

$$k \sim \alpha \log n, \tag{15}$$
$$R \sim \alpha \log \left( 1 + \frac{1}{\alpha} \right) \log n,$$

for some $\alpha > 0$. These values of $k$ and $R$ yield to the rate-per-link

$$\lambda = \log \left( 1 + \frac{1}{\alpha} \right), \tag{16}$$

which is a constant. In fact, this increased rate-per-link is the result of allowing less users to be simultaneously active.

A closer examination of (15) and (16) reveals a tradeoff between the number of active users and the rate-per-link. To formulate this tradeoff let us define

$$\kappa = \lim_{n \to \infty} \frac{k_n}{\log n}, \tag{17}$$

to be the scaling factor of the number of users. Then, from this definition and from (15) and (16) we obtain

$$\lambda = \log \left( 1 + \frac{1}{\kappa} \right). \tag{18}$$

### B. Non-Zero Rate-per-Links by Centralized On-Off Strategy

According to Lemma 2, when $k \sim n^{\alpha+\epsilon_n}$ for some $0 < \alpha < 1$ and some $\epsilon_n \to 0$, the sum-rate scales as $(1-\alpha) \log n$, but the rate-per-link approaches zero as $n \to \infty$. However, if a centralized link activation is allowed, it will be possible to choose a subset of the links with small mutual interference such that the rate of each link is in the order of $\Theta(1)$. The selection process should be such that the number of active links is large enough to avoid deteriorating the network throughput. As will be seen later, this is guaranteed since the

---

$$R \sim \begin{cases} o(\log n) & \text{if } k = o(\log n) \\ \alpha \log \left( 1 + \frac{1}{\alpha} \right) \log n & \text{if } k \sim \alpha \log n \text{ for some } \alpha > 0 \\ \log n & \text{if } k = \omega(\log n) \text{ and } k = o(n^\alpha), \quad \forall \, 0 < \alpha < 1 \\ (1-\alpha) \log n & \text{if } k \sim n^{\alpha+\epsilon_n} \text{ for some } 0 < \alpha < 1 \text{ and some } \epsilon_n \to 0 \\ o(\log n) & \text{if } k \sim n^{1-\epsilon_n} \text{ for some } \epsilon_n \to 0^+ \end{cases} \tag{14}$$

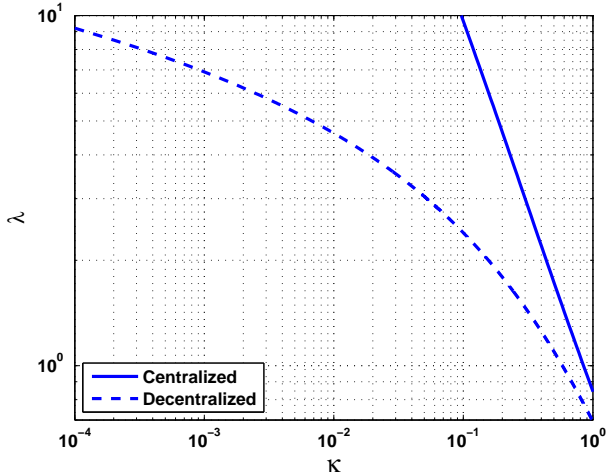

Fig. 1. Tradeoff between rate-per-link and the number of active links.

number of primarily selected links, i.e. $k$, is large. Strategy 2 describes one possible interference management policy.

*Strategy 2:* For a threshold $\delta > 0$, from the set of links whose direct channel coefficients are larger than $t = (1-\alpha)\log n$, choose the links whose cross channel coefficients are less than $\delta$.

Since Strategy 2 requires the knowledge of all channel coefficients by a central unit, it can only be implemented as a centralized scheme. In the following, we show that one can choose the value of $\delta$ optimally such that the sum-rate is maximized; this optimum value yields to rate-per-links of order $\Theta(1)$.

Assume $\boldsymbol{V} = \{1, \cdots, k\}$ is the set of links whose direct channel coefficients are larger than $t = (1-\alpha)\log n$. Construct an undirected graph $G(\boldsymbol{V}, \boldsymbol{E})$ with vertex set $\boldsymbol{V}$ and the adjacency matrix $\boldsymbol{E} = [e_{ij}]$ defined as

$$e_{ij} = \begin{cases} 1 & ; \quad g_{ij} \leq \delta \text{ and } g_{ji} \leq \delta \\ 0 & ; \quad \text{otherwise} \end{cases}.$$

It can be easily verified that the probability of having an edge between vertices $i$ and $j$ for the special case of Rayleigh fading channel equals

$$\pi = \left(1 - e^{-\delta}\right)^2. \tag{19}$$

The definition of $G$ implies that $G \in \mathcal{G}(k, \pi)$, where $\mathcal{G}(k, \pi)$ is the family of $k$-vertex *random graphs* with edge probability $\pi$.

In strategy 2, we are interested to choose the maximum number of links whose interference coefficients are smaller than $\delta$. This is equivalent to choose the largest complete subgraph of $G$. The size of the largest complete subgraph of $G$ is called its clique number. The following lemma provides the scaling law of the clique number of a random graph.

*Lemma 3 ( [13]):* For a fixed $0 < \pi < 1$, the clique number $X_k(G)$ of $G \in \mathcal{G}(k, \pi)$ scales as

$$X_k \sim \frac{2 \log k}{\log 1/\pi} \qquad a.a.s.$$

Consequently, by using $\pi$ from (19) and with $k \sim n^{\alpha+\epsilon_n}$, the number of links selected by Strategy 2 is obtained as

$$\hat{k} \sim \frac{-\alpha}{\log\left(1 - e^{-\delta}\right)} \log n. \tag{20}$$

With this number of users and by taking the same approach as in the derivation of (9), the sum-rate is lower bounded as

$$R \geq \frac{-\alpha}{\log\left(1 - e^{-\delta}\right)} \log \left(1 - \frac{(1-\alpha)\log\left(1 - e^{-\delta}\right)}{\alpha\left(1 - \dfrac{\delta e^{-\delta}}{1 - e^{-\delta}}\right)}\right) \log n. \tag{21}$$

Note that $(1 - \dfrac{\delta e^{-\delta}}{1 - e^{-\delta}})$ in the denominator is the expected value of the cross channel coefficients conditioned on being less than $\delta$.

The lower bound in (21) is logarithmic in $n$ as desired. Since the number of active links in (20) is logarithmic in $n$ as well, we conclude that rate-per-link scales as $\Theta(1)$.

The scaling factor of $\log n$ in (21), which is independent of $n$, can be numerically maximized with respect to $\delta$. Similar to (18) that describes the tradeoff between $\lambda$ and $\kappa$ (the scaling factor of the number of active links defined in (17)) for the centralized scheme, here we can find this tradeoff numerically. Fig. 1 demonstrates the tradeoff between $\lambda$ and $\kappa$ for the centralized and decentralized schemes. For the centralized scheme the value of $\delta$ that maximizes (21) has been used. As observed, for a ceratin value of $\lambda$ the centralized scheme can support larger number of users, especially for larger values of $\lambda$. It should be mentioned that the decentralized scheme not only achieves non-zero rate-per-link, but provides a sum-rate close to $\log n$ at large values of $\lambda$.